\documentclass{article}
\usepackage[utf8]{inputenc}
\usepackage{authblk}
\usepackage{setspace}
\usepackage[margin=1.1in]{geometry}
\usepackage{graphicx}
\usepackage{tabularx}
\usepackage{subfigure}
\usepackage{amsmath,soul}
\usepackage{lineno,xcolor}
\usepackage{lipsum}

\usepackage[style=ieee, 
citestyle=numeric-comp,
sorting=none]{biblatex}
\newcommand{\Tesla}{\mathrm{T}}
\newcommand{\Hz}{\mathrm{Hz}}

\title{Quantum sensing on magnetic field inspired by Avian compass}

\author[1]{Wei-Yin Chiang}
\author[2]{Yuan-Chung Cheng}
\author[1]{Min-Hsiu Hsieh}
\affil[1]{Hon-Hai Quantum Computing Research Center, Taipei, Taiwan }
\affil[2]{Department of Chemistry, National Taiwan University, Taipei, Taiwan}

\date{}
\begin{document}
\maketitle
 \begin{abstract}

Magnetic measurement can be performed by various sensors, such as SQUID and Giant Magnetoresistance. This device can achieve high accuracy while losing efficiency and convenience. The model of biological magnetic sensing in avian proposes a radical pair response to the external field on the FLY results in regulating animal behaviour. Inspired by the radical pair system found in biological system, the effect of intra-radical coupling and the initial condition is studied in this simplified radical pair model as the quantum advantage results from entanglement and superposition are investigated in metrology. To identify the sensing benefit from the cooperation between the radical pairs, the inter-radical coupling is considered. We find the intra-radical coupling determines the sensing regime while the coupled radical pair system enables a new sensing scheme that provides a more general and flexible way for magnetic sensing in a broader range through a easier manipulation to the system. \footnote{Provisional patent application has been submitted}
 
\end{abstract} 
\section{Introduction}

In quantum metrology, magnetic field is measured by asserting the spin dynamics of a quantum system regulated by the magnetic field, known as Ramesy measurement \cite{Giovannetti:2011}. Utilizing the advantage of superposition and entanglement in quantum systems, good sensitivity and accurate measurement can be achieved beyond that of a typical classical one. More specifically, by applying operations to initial entangled states, the phase of the system state is amplified and the Heisenberg limit could be achieved with the uncertainty of $1/N$, where $N$ is the number of qubit, instead of $1/\sqrt{N}$ for a typical classical measurement.

Highly sensitive, precise and flexible magnetic field measurement at nano-scale is of fundamental importance to spintronic and biological applications \cite{Schirhagl:2014}. The ultra sensitive measurements typically utilize, for example, spin-dependent scattering in Semiconductor Magnetoresistive Element and Giant Magnetoresistance (GMR) \cite{Binasch:1989},  change of Josephson phase in superconducting quantum interference device (SQUID) \cite{Jaklevic:1964} and Zeeman effect in Nitrogen-vacancy (NV) spin defect \cite{Rondin:2014}. The fascinating sensitivity of the above mechanisms can be $\sim 10^{-15}\Tesla/\sqrt{\Hz}$, however, all those measurements except NV defect can not be performed under ambient pressure at room temperature.

NV defect spin system senses the magnetic field through utilizing the Zeeman effect of the unpaired electron induced by Nitrogen-vacancy defects \cite{Rondin:2014}. This quantum sensing system can fulfill strong demands on spatial resolution in applications, such as measuring the induced field from nanoscale chips \cite{Appel:2015}, magnetite crystal from magnetotactic bacteria \cite{Schirhagl:2014}\cite{McCoey:2020}\cite{Gille:2021}. The information of magnetic field is read out by electron spin resonance (ESR) and Ramsey measurement with the sensitivity of $10^{-6} \Tesla/\sqrt{\Hz}$ and $10^{-9} \Tesla/\sqrt{\Hz}$, respectively. The spatial resolution ($\sim$ 10-100 nm) of ensemble NV${^-}$ sensing is limited by the density of defects (in $\sim$ ppm) and the size of the crystal, and hence is limited by its fabrication method. The point defects in NV-center are artificially implemented by ion gun \cite{Pezzagna_:2010} or detonation of explosives in a closed vessel \cite{Kuroyama:1967}. It is hard to control the number, positions and orientation of defects precisely. One has to make the crystal smaller to achieve better spatial resolution, however, the signal intensity is reduced as well, resulting into a scarification of measurement precision. Moreover, the positions of defects are still fixed and lack flexibility. As a result, NV defect spin system lacks a strong flexibility to be easily tuned to have a balance of spatial resolution and precision requirement, to adapt to different magnetic field measurements.

In this manuscript, we will demonstrate a novel magnetic field sensing system based on intra/inter-radical coupling in radical pair system. By adjusting the intra/inter-radical coupling, one achieves a more general and flexible sensing scheme, fullfiling both the requirement of adjustable resolution and precision. In fact, radical paring magnetic sensing has a deep root in biological system. The magnetic field from the earth provides geological information to migratory birds for long-distance precise travel \cite{Wiltschko:2019}. Molecule radical sensing is proposed to be the potential biological mechanism of this avian compass \cite{Hore:2016}\cite{Schulten:1978}.%
According to the model proposed by Peter Hore, \emph{Cryptochrom}(\emph{Cry}) is the signal protein that attaches to the membrane of rod cell located at the retina and is responsible for magnetic field sensing. Flavin adenine dinucleotide (FAD) cofactor binds on the N-terminal photolyase domain of \emph{Cry}, and the electron transfer from tryptophans-400 (Trp-400) to FADH$^+$ is induced by the blue light. The electric hole resulted from the electron transfer is therefore propagated through Trp-377 to Trp-324, and the process sustains about 300ns. This electron spin system has four states: three triplet states and one singlet state. The yields of the singlet/triplet state are accompanied by the conformation change that activates the signalling cascade. This signal amplification process regulates neuron firing rate by activating ion channels, so that birds can sense the earth's weak magnetic field ($\sim 50\times 10^{-6} T$) and be guided precisely to their destination in their long journey.

The sensing model had been verified by the experiment \cite{Kerpal:2019} on a molecule radical pair system consisting of carotenoid-porphyrin and fullerene moieties. Both the angle and magnitude of the magnetic field can be detected by molecule radical pair. Numerous theoretical and numerical studies have also examined the sensing mechanisms;\cite{Hore:1998A, Hore:1998B} found that the recombination rate  plays a critical role in magnetic sensing; the exchange and dipolar interaction in radical pair \cite{Hore:2008} is partially cancelled in order to have an efficient inter-conversion between singlet and triplet states; a multi-step radical transfer process will enhance the sensitivity \cite{Hore:2016, Wong:2021A, Wong:2021B}; and hyperfine interaction from the real molecular structure allows precise magnetic direction sensing \cite{Hiscock:2016}. Those studies focused on exploring the possibility of the model of protein radical pair as the still debating biological magnetic sensing mechanism.

Here we push this radical pair model system into a more general framework, by including the intra/inter-radical-pair coupling as a tuning blob, we demonstrate that such a tuning can lead to a quantum magnetic sensing system which is both flexible and precise, overcoming the limitation of NV defect system. The sensing is performed under ambient pressure and room temperature, which can not be achieved by SQUID, GMR or Hall effect. In general, intra-radical coupling can unnecessarily be partially cancelled in an arbitrary molecule radical pair system. By introducing the intra-radical coupling $G_{AB}$ to the model system, the working regime for magnetic sensing changes with this $G_{AB}$. This allows one to obtain an accurate measurement of magnetic field by choosing an appropriate intra-radical coupling. Modification on intra-radical coupling can be achieved by changing the molecule radical pair. To have a flexible measuring scheme, we propose a sensing method in the following discussion. Collective sensing usually results in higher sensitivity and robustness in biological sensing \cite{Prentice:2016}. The sensing regime can be changed by adjusting the coupling strength while the intrinsic property of the cell is not varied too much. Inspired by this evidence, the radical sensing system is studied by introducing weak coupling between radical pairs. Response pattern shows that the yield production rate varies with the inter-radical coupling. Theoretical analysis shows the energy degeneracy results from inter-radical coupling and the Zeeman effect from the external field, allowing us to deduce the field strength from the peak production. Therefore, a new sensing scheme that is convenient without losing flexibility is proposed. One can sense the broad range of magnetic fields by only one type of radical pair.



\subsection{Overview of Main Results}

Inspired by radical pair sensing to the magnetic field in the biological system, in this manuscript, we demonstrate the peak of singlet yield is found at a specific magnetic field. This peak shifts to the stronger magnetic field when the intra-radical coupling is increased and forms a V-shape pattern. This pattern is realized by energy degeneracy under the suitable magnetic field and intra-radical coupling, allowing one to obtain the required sensing regime by adjusting intra-radical coupling in an isolated radical pair system.  
A more general and flexible sensing scheme is proposed by a coupled-radical pair system. Unlike the response pattern found in an isolated radical pair, a more complicated pattern is observed in a two-coupled radical pair system. More than one yield peak can be found under the constant magnetic field when the inter-radical coupling is increased. This phenomenon can be utilized for a new method of magnetic sensing as the mechanism can be explained by the first-order approximation. According to the approximation, the system energy degeneracy will be happening at some energy gaps under the specific inter-radical coupling results in the yield peaks. From this analysis, the magnetic field can be deducted from the production peaks. Therefore, a general and flexible sensing method can be easily achieved by screening the distances between radical pairs that in terms of changes in coupling strength.

Besides the main findings mentioned above, we find (1)the coupling configuration that allows one to obtain the sensitive response is $\hat{S}_B$-$\hat{S}_B$ coupling; (2)local interaction determines the response pattern; (3)that better sensitivity can be achieved when the sensing system starts with a quantum interference state.

\subsection{Related Works}

We compare the state-of-the-art magnetic sensing system by looking at system fabrication, the sensing mechanism, system initialization, signal readout, sensitivity, spatial resolution, working environment and generality. Here we consider quantum sensing systems only.  From the table, we find the existing system can have good sensitivity while the spatial resolution is restricted.


\begin{center}
\begin{tabular}{ |c|c|c|c| }

 \hline
   & SQUID\cite{Fagaly:2006} & NV$^{-}$-defect\cite{Rondin:2014}\cite{Taylor:2008} & Coupled radical pair \\
 \hline
 Fabrication & Superconducting loop  & Controlled detonation crystal   & Artificial synthesis molecule  \\
 Mechanism & Josephson phase change  & Zeeman effect  & Zeeman effect \\
 System initialization & - & Laser pumping $\lambda =532nm$ & blue light $\lambda = 450 nm$ \\
 Signal readout & current & Photon sensing $\lambda =637 nm$ & Signal absorption \\
 Observation & I-V curve  & Electron spin resonance(ESR) & Singlet yield production \\
 Sensitivity$^a$ &  $1\mathrm{p T/\sqrt{Hz}}$ & $1 \mathrm{\mu T/\sqrt{Hz}}$ & $1 \mathrm{nT/\sqrt{Hz}}^d$ \\
 spatial resolution & $1\mathrm{\mu m}$ & 10-100nm  & $<$ 10 nm  \\
 Working environment & 77K & S.T.P.$^c$ & S.T.P \\
 Generality$^b$ & None & None & High \\
 
\hline
\end{tabular}
\end{center}

Comparison of magnetic sensing methods in quantum sensing system.\\
a. Sensitivity: $\eta$, $B_{min}\sqrt{T}$ , where $B_{min}$ is the minimum detectable magnetic field and $T$ is measurement duration.\\
b. Generality is defined by whether the distance between sensing elements is tunable.\\
c. S.T.P.: Standard temperature and pressure\\
d. Sensitivity is estimated through the definition in Eq.\ref{sensitivity}. Note that the scale is corrected by assuming the hyperfine interaction is 
$\mathrm{0.01 mT}$.


\section{Model}
The Hamiltonian of radical pair consists of hyperfine interaction, dipole-dipole interaction, exchange energy and Zeeman effect. Here we aim to understand the fundamental mechanism of this quantum magnetic sensing in radical pair system, the Hamiltonian is reduced to the description of one nucleus and two electrons. This simplified Hamiltonian is written as:\\
\begin{align}
     H = a \hat{I}\cdot \hat{S}_A + G_{AB}\hat{S}_A^z\cdot \hat{S}_B^z +\theta (\hat{S}_A^z + \hat{S}_B^z)
\end{align}

Where $\theta$ is the magnetic field that we aim to detect, $a$ is the parameter controls hyperfine interaction, only spin $\hat{S}_A$ interact with nucleus $\hat{I}$ while magnetic field acts on both spin $\hat{S}_A$ and $\hat{S}_B$ in $z-$direction. $G_{AB}$ is the intra-radical coupling that controls spin-spin interaction between two electron. Given the Hamiltonian, the dynamics of system can be expressed by $\rho(t) = e^{iHt}\rho_0 e^{-iHt}$ and $\rho_0$ is the density matrix at $t=0$. Singlet yield, $y_s(t)$ , can be obtained by projecting the singlet state onto the density matrix, $\mathrm{Tr}[ \hat{P}^s  \rho(t)]$. The singlet yields can be projected to the eigenstate of Hamiltonian, 
$$ y_s(t) = \frac{1}{M} \sum_{m=1}^{4M} \sum_{n=1}^{4M} \langle m| \hat{P}^s |n\rangle\langle m| \rho _0 |n\rangle cos(\omega_{mn})$$
$M$ is the number of nucleus spin configuration.

\begin{figure}[h!]
\centering
\includegraphics[width=0.5\textwidth]{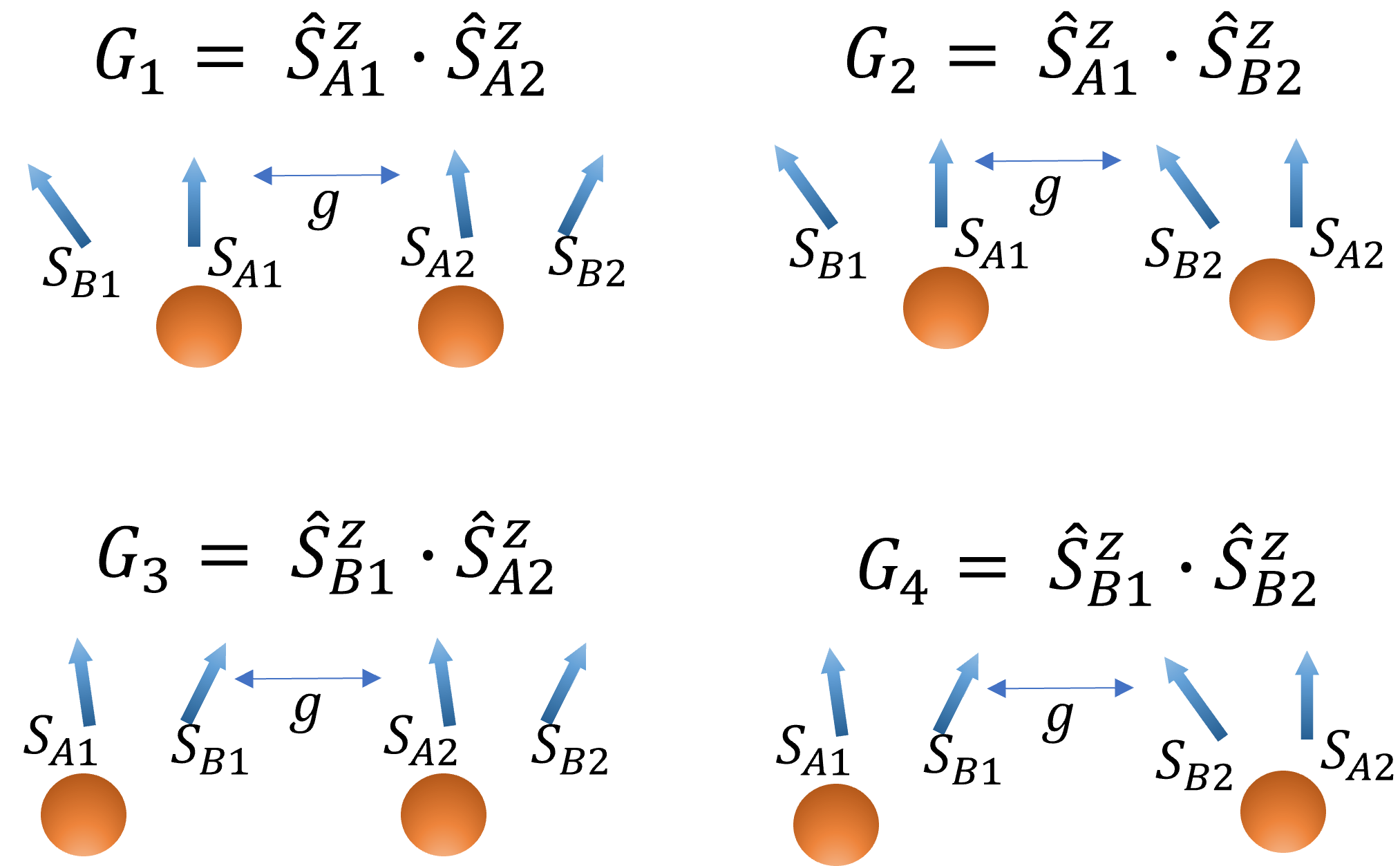}
\caption{Schematic expression of four possible coupling configurations.}\label{FIG:2-coupled_all}
\end{figure}

In the proposed sensing model, the signal level is accumulated during the life time of radical pair. The total singlet yields is
\begin{align}
\label{totaly}
\Phi_S(\theta) = k \int_{0}^{\infty} y_s(t) e^{-kt} dt = \frac{1}{M} \sum_{m=1}^{4M} \sum_{n=1}^{4M}\hat{P}^s_{mn}\rho^0_{mn} f(\omega_{mn})
\end{align}
Here $\hat{O}_{mn} = \langle m |\hat{O} |n \rangle$, $f(\omega_{mn}) =\frac{k^2}{k^2+ \omega_{mn}^{2}}$. 
From the expression, it is easy to see the information of the magnetic strength has been enclosed in the Hamiltonian, therefore, eigenvalues and eigenvectors. Since the singlet yields changes with the magnetic field, by measuring the total singlet yield $\Phi_s$, one can find the magnetic field. The relation between the magnetic field and the amount of singlet yield gives the response curve that shows how singlet yield changes with the magnetic field. This curve illustrates the singlet yield changes by slightly changing the magnetic field. On the other hand, sensitivity describes the minimum magnetic field can be sensed within the square root of measuring duration and is equivalent to the inverse of response level of the system to external stimuli under the current adapted state.The sensitivity $S$ is defined\cite{Rondin:2014} by the ratio of $\sqrt{\Phi_s}$ and $\partial_{\theta} \Phi_s$.

\begin{align}
\label{sensitivity}
 S = \sqrt{\Phi_s}/ \frac{\partial \Phi_s}{\partial \theta}
\end{align}
For coupled radical pair system, Hamiltonian becomes:
\begin{equation}
    H = a \sum_{i=1}^{n} \hat{I}_i\cdot \hat{S}_{Ai} + G_{AB}\hat{S}_{Ai}^z\cdot \hat{S}_{Bi}^z + \theta (\hat{S}_{Ai}^z + \hat{S}_{Bi}^z)+g \sum_{ij}G(\hat{S}_{Ai}^z,\hat{S}_{Bi}^z,\hat{S}_{Aj}^z,\hat{S}_{Bj}^z)
\end{equation}

where $G$ is inter-radical coupling function that describes spin-spin interaction, $g$ is the inter-radical coupling strength, $G_{AB}$ is the intra-radical coupling strength. Here we set $a=1$. In a two-coupled radical pair system, one can find four possible coupling methods and are shown in Fig.\ref{FIG:2-coupled_all}.

Since $G_2$ and $G_3$ corresponds to the same configuration, only three coupling methods for a two-coupled pair system are considered.

\section{Results}

\subsection{Sensitive regime changes with the interaction within a radical pair}

According to the results from Peter Hore\cite{Hore:1998A}, the abrupt change in the singlet yield production allows the sensing at the weak field regime. We wonder if the working regime of the field sensing will be different when the interaction between radicals is considered. Considering the situation with $G_{AB} \neq 0$, we found the responses of a radical pair to the external field change with $G_{AB}$. Under the initial condition of singlet state which is generated by nature, the singlet yields is in the form of\\
\begin{multline}
\label{eqn:singlet_GAB}
    \Phi_s = \frac{1}{4}+\frac{1}{8}[\frac{a^4}{(a^2+\gamma_1^2)^2}+\frac{a^4}{(a^2+\gamma_2^2)^2}+\frac{\gamma_3^4}{(a^2+\gamma_3^2)^2}+\frac{\gamma_4^4}{(a^2+\gamma_4^2)^2}]+
     \frac{a^2}{4(a^2+\gamma_2^2)}f(\frac{a}{2}-\frac{\gamma_2}{2}) \\ +\frac{a^2}{4(a^2+\gamma_1^2)}f(\frac{a}{2}-\frac{\gamma_1}{2}) +
    \frac{\gamma_3^2}{4(a^2+\gamma_3^2)^2}f(\frac{a}{2}-\frac{\gamma_4}{2}) + \frac{\gamma_3^2}{4(a^2+\gamma_4^2)^2}f(\frac{a}{2}-\frac{\gamma_3}{2})  \\
     +\frac{a^4}{4(a^2+\gamma_2^2)(a^2+\gamma_1^2)}f(p)+\frac{\gamma_3^2\gamma_4^2}{4(a^2+\gamma_3^2)(a^2+\gamma_4^2)}f(o)
\end{multline}

where $\gamma_1 = 2G_{AB}+p-\theta$, $\gamma_2 = 2G_{AB}-p-\theta$, $\gamma_3 = 2G_{AB}+o+\theta$, $\gamma_4 = 2G_{AB}-o+\theta$, $o=\sqrt{a^2+(\theta+2G_{AB})^2}$ and $p=\sqrt{a^2+(\theta-2G_{AB})^2}$. 

\begin{figure}[h!]
\centering
\includegraphics[width=0.5\textwidth]{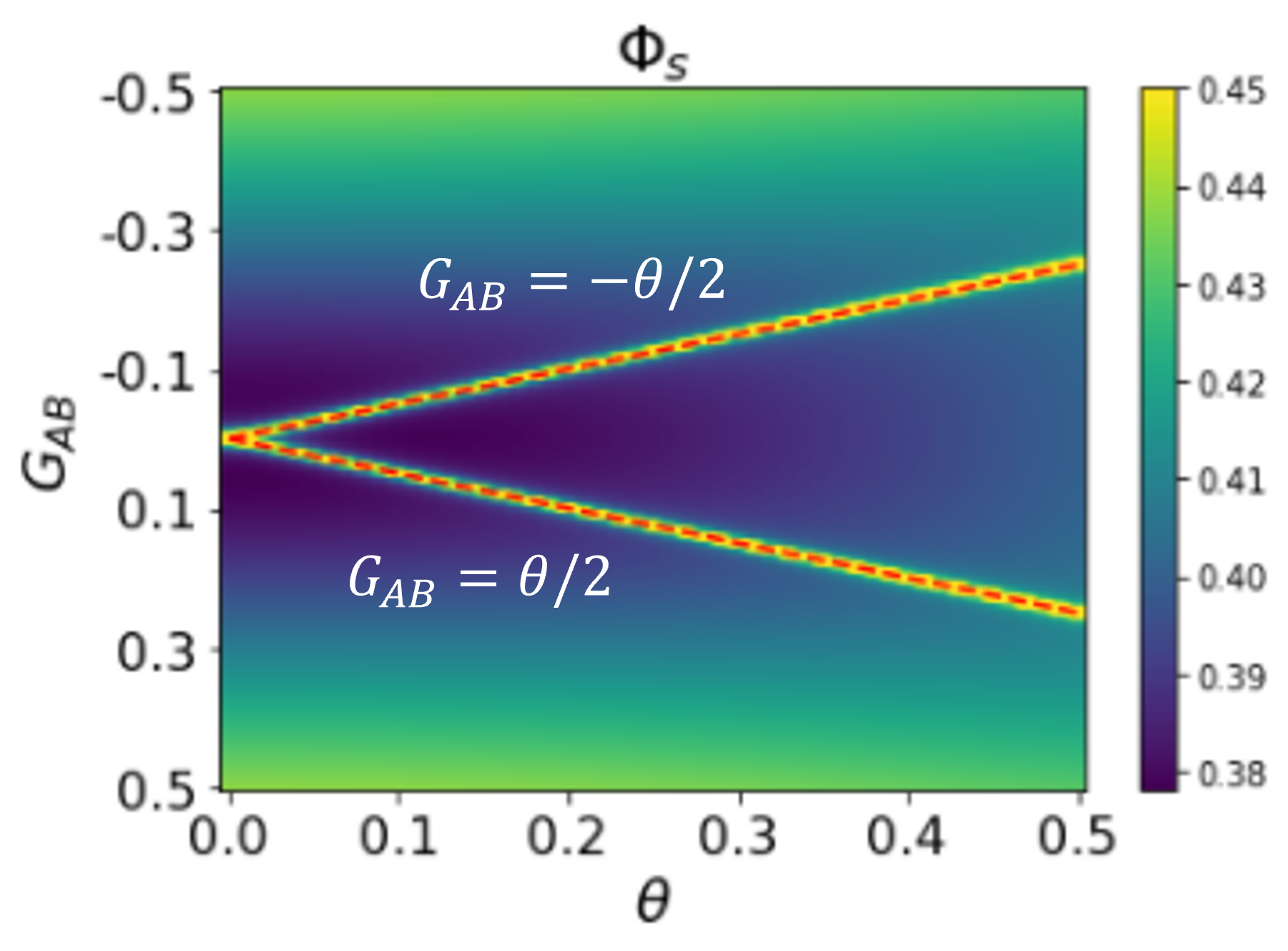}
\caption{Contour plot of singlet yield under different spin-spin interaction and magnetic field. Under a specific $G_{AB}$, one can find a peak from the response curve. The peak moves with $G_{AB}$ and can be described by the red dot line(color online) $G_{AB}=\pm \frac{\theta}{2}$.}\label{FIG:1_radical_GAB}
\end{figure}

One can see how the singlet yield is varied with the magnetic field by changing the external field under the constant $G_{AB}$ from the response curve.  Concatenating the response curves under different $G_{AB}$ gives the response pattern of singlet yield.
From the response pattern in Fig.\ref{FIG:1_radical_GAB}, the V-shape pattern is formed by yield peaks can be explained by Eq.\ref{eqn:singlet_GAB}. From Eq.\ref{eqn:singlet_GAB} we can see the maximal yield can be found when $p=0$ or $o=0$. These two conditions give the red dash line in Fig.\ref{FIG:1_radical_GAB} that $G_{AB} = \frac{\theta}{2}$. 

The sensitive regime can be identified when the small amount of change in external field results in the strong change in yield production. From Fig.\ref{FIG:1_radical_GAB}, we found each $G_{AB}$ can be sensitive at different magnetic regimes. This property allows us to quantify the magnetic field accurately by choosing $G_{AB}$ smartly. It is a possible evolutional evidence that the bird adjusts the intra-radical distance to optimize the sensitivity base on the the weak environment signal cue from the earth. Therefore, the angle of magnetic field can be deducted from the the magnitude of the magnetic field in different directions. For molecule radical system, the different intra-radical distance can be achieved by synthesis of biradical molecules. However, from the results shown in appendix, we found the response pattern with $G_{AB} \ne 0$ shows the behaviour qualitatively similar to that of $G_{AB} = 0$ while it had been demonstrated that the cancellation between dipole-dipole interaction and exchange energy for the spin-spin separation $~1.9$ nm in \emph{Cryptochrom}, in the following content, only $G_{AB}=0$ is considered . 

\subsection{Coupling enables new sensing scheme}

\begin{figure}[h!]
\includegraphics[width=1\textwidth]{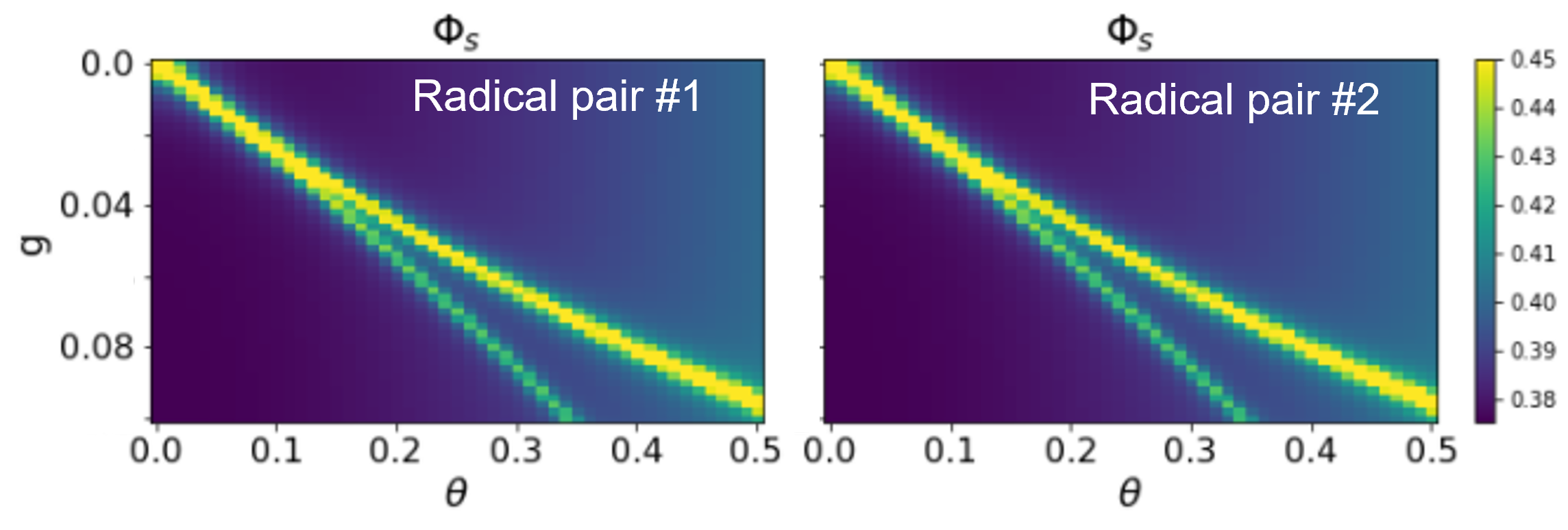}
\caption{ Coupling between $S_{B1}$ and $S_{B2}$($G_4$ coupling) is considered. Contour plot of singlet yield in both radical pairs. Because of the coupling method, the configuration is symmetric, both radical pair \#1 and \#2 show the identical response to the external field under coupling.}\label{FIG:3-RP_couple_1}
\end{figure}

Although the isolated radical pair can have different working regimes by tuning $G_{AB}$. Adjusting the intra-radical distance on the fly is not easy, a more general sensing method that can be applied to a broader sensing regime in a flexible way is needed during measurement. In the biological system, the sensing process usually involves collective behaviour. Coupling between the cells is the most frequent path to propagate information. Here we investigate the effect of inter-radical coupling on the sensing behaviour. Starting from two-coupled radical pair, we consider the configuration shown in Fig.\ref{FIG:3-RP_couple_1}(a), that the coupling is introduced between $S_{B_1}$ and $S_{B_2}$ ($G_4$ coupling), the contour plot in Fig.\ref{FIG:3-RP_couple_1}(b) illustrates how the singlet yield changes with coupling strength($g$) and magnetic field. We find that increases in coupling strength not only move the position of the singlet yield peak to a higher magnetic regime but also split it. These hilly regimes give the sensitive response. Coupling not only enlarges the sensitive regime but allows us to design the new sensing scheme.  

To design a new sensing scheme, we study the mechanism of the peak formation. Introducing coupling between radical pairs brings different sensing scenario. Coupling gives a complicated energy spacing results in the different singlet yield production under the same magnetic field. The singlet yield peaks in a coupled radical pair system are contributed by the resonance between the coupling strength and the energy spacing. This can be explained by performing the first-order approximation to the system. For simplicity, consider a two-coupled radical pair system under the weak coupling, takes the coupling term to be a perturbation term. Its eigenvectors can be obtained from the unperturbed Hamiltonian while eigenvalues need to be corrected by adding the perturbation term to the unperturbed eigenvectors. This corrected energy gapes explains the mechanism of peak formation. 
To explain in detail; for single radical pair, the energy gapes and the corresponding eigenmodes can be obtained by finding the eigenvalue $\lambda_1$ and eigenvectors $|m_1 \rangle$ from the Hamiltonian of the isolated radical pair, $H_1$. When the eigenvalues and eigenvectors of the second isolated radical pair Hamiltonian, $H_2$, are denoted by $\lambda_2$ and $|m_2\rangle$, the total Hamiltonian of two-coupled radical pair system is $H_1 \otimes I_8 + I_8 \otimes H_2$. The corresponding eigenvalues and eigenvectors become $\lambda_1+\lambda_2$ and $|m_1\rangle \otimes |m_2\rangle$. When the coupling term is considered, according to the the first order approximation, the eigenvectors remain unchanged when the eigenvalues becomes $\lambda_1+\lambda_2+ g\langle m_1 m_2|G|m_1 m_2\rangle$. Remind that the singlet yield $\Phi_s$ is 
$\sum_{i,j}\sum_{m_{ij},n_{ij}}|\hat{P}_{m_{ij} n_{ij}}^s|^2\frac{k^2}{k^2+\omega^2_{m_{ij} n_{ij}}}$. 
Appropriate parameter($\theta$ and $g$) combinations result in peak formation. In this situation, we will have large $|\hat{P}_{m_{ij} n_{ij}}^s|$ and small $\omega^2_{m_{ij} n_{ij}}$. For example, under coupling $G_4$, the energy gaps corresponding to the dominated components(large $|\hat{P}_{m_{ij} n_{ij}}^s|$ ), $m_{ij}$ and $n_{ij}$, are:\\
$$2g+\frac{\Omega}{2}-\frac{1}{2}(1+\theta), 2g +\frac{\Omega}{2}-\frac{1}{2}(1-\theta)$$
$$2g +\frac{\Omega}{2}-\frac{1}{2}(1-\theta), 2g - \frac{\Omega}{2}-\frac{1}{2}(1+\theta)$$
$$2g +\frac{\Omega}{2}-\frac{1}{2}(1+\theta), 2g - \frac{\Omega}{2}-\frac{1}{2}(1-\theta)$$
$$2g - \frac{\Omega}{2}-\frac{1}{2}(1-\theta), 2g - \frac{\Omega}{2}-\frac{1}{2}(1+\theta)$$
Where $\Omega=\sqrt{1+\theta^2}$. When the degeneracy happen, peaks appear. However, within the eight energy gaps, only last four give positive coupling strengths. They are:
$$g=\frac{1}{4}[\theta\pm (1-\Omega)], g=\frac{1}{4}[(1+\Omega)\pm \theta]$$
The first two successfully predict the peak positions while the last two give the strong coupling strength that perturbation theory may fail. One can find the peaks under the constant magnetic field at $g_1$ and $g_2$, from the expression and deduce the magnetic field from these two peak positions by \\
$$g_1+g_2 = \frac{1}{2} \theta \Longrightarrow \theta=2(g_1+g_2) $$
The expression above allows one to measure the magnetic field by changing the radical pair distance that relates to the changes in coupling strength. Since the modification in distance will induce the peaks of singlet yield, the magnetic field can be estimated from the converted coupling which is modulated by the distance between the radical pairs.

\subsection{The sensitivity of $\hat{S}_B$-$\hat{S}_B$ coupling system outperforms other coupling configurations  }

Sensitive regime can be adjusted by inter-radical coupling. From the observation, the sensitive regime is pushed to the weaker magnetic regime when the connection is established at $S_A$. In a two-coupled radical pair system, the structure of the radical pair is not symmetric, different coupling configurations, $G_1$ and $G_2$ are considered further.

\begin{figure}[h!]
\centering
\includegraphics[width=1\textwidth]{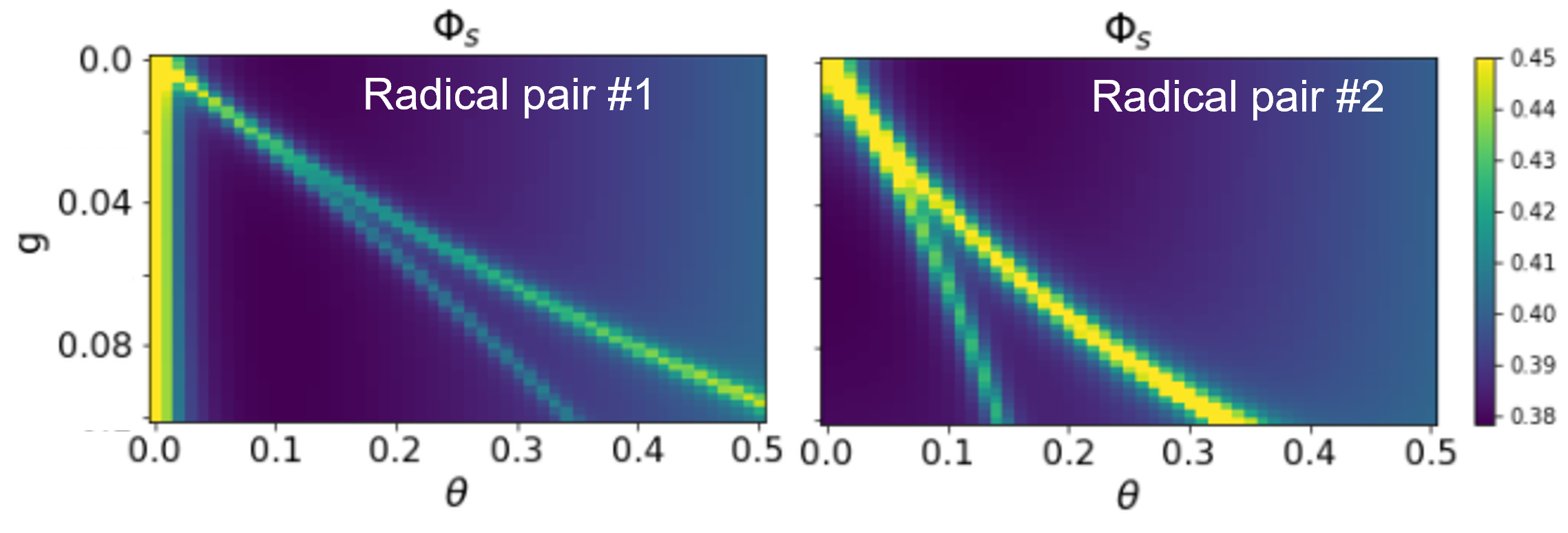}
\caption{Coupling between $S_{B1}$ and $S_{A2}$($G_2$ coupling) is considered. Contour plot of singlet yield in both radical pairs. Radical pair \#1 shows the weaker response compare to $G_4$, while the response regime of radical pair \#2 is squeezed to the weaker magnetic regime. Since the coupling configuration is symmetric, both radical pair \#1 and \#2 show the identical response to the external field under coupling. 
}\label{FIG:4-RP_couple_23_0}
\end{figure}

\begin{figure}[h!]
\centering
\includegraphics[width=1\textwidth]{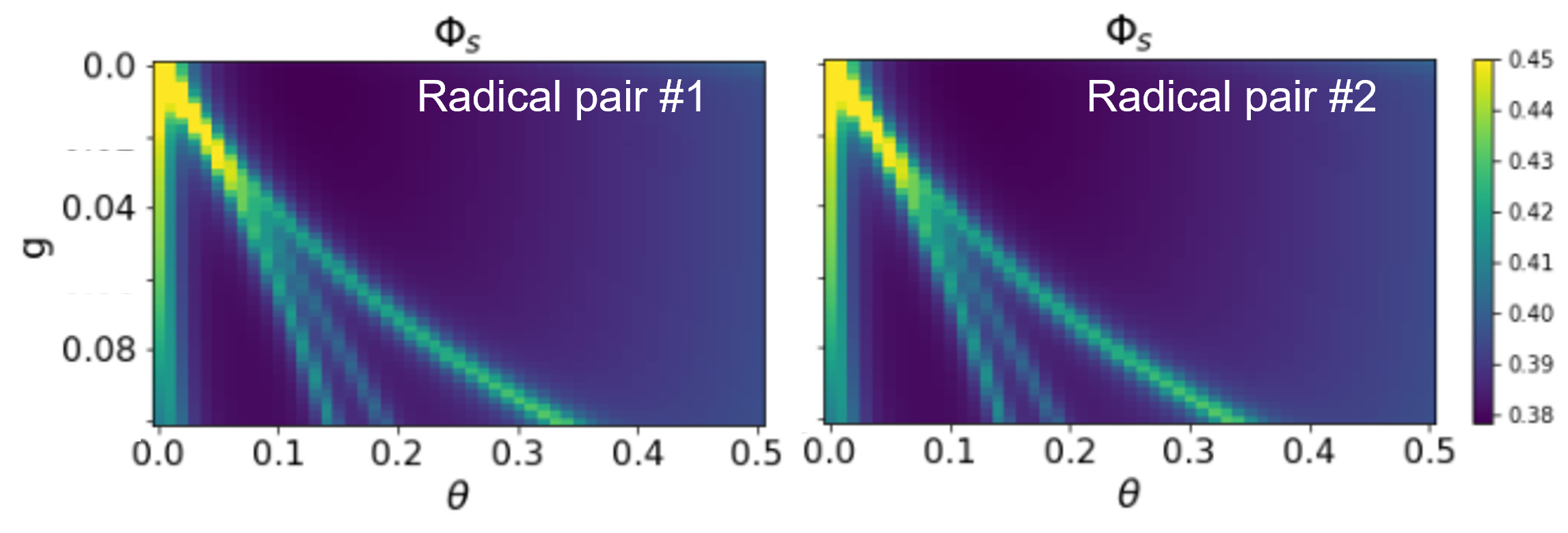}
\caption{Coupling between $S_{A1}$ and $S_{A2}$($G_1$ coupling) is considered. Contour plot of singlet yield in both radical pairs. Because of the coupling method, the configuration is symmetric, both radical pair \#1 and \#2 show the identical response to the external field under coupling. Under the $G_1$, not only the peak splitting appears in small magnetic regime, but the response regime is squeezed.
}\label{FIG:4-RP_couple_23_1}
\end{figure}

Coupled radical pair under $G_2$ in Fig.\ref{FIG:4-RP_couple_23_0} (a) gives the response patterns of each radical pair in Fig.\ref{FIG:4-RP_couple_23_0} (b). Different response patterns can be found at each radical pair. If we compare the response patterns under $G_2$ and $G_4$ coupling, a weaker response at radical pair \#1 and a smaller sensing regime at radical pair \#2 under $G_2$ coupling are observed. The weaker signal at radical pair \#1 can be realized by inspecting the energy gapes corresponding to the dominated components modified by perturbation term under the approximation in the previous section. By performing the first-order approximation to the systems with $G_2$ and $G_4$ coupling, respectively, the perturbation term determined by coupling function takes part in the energy correction term, therefore energy gaps. $G_2$ and $G_4$ share the same dominated components with different energy gaps while the coupling configuration of $G_2$ is different from $G_4$ by coupling to $S_{A2}$. The weaker response in radical pair \#1 is attributed to the less degeneracy components with eligible $g$ under $G_2$ coupling configuration. Among all coupling methods, $G_4$ enables the system to sense the broader range of magnetic field with a stronger response signal compared to the system under $G_2$. For the system with the coupling configuration $G_1$ in Fig.\ref{FIG:4-RP_couple_23_1}(a), the response behaviour can be even more complicated. From Fig.\ref{FIG:4-RP_couple_23_1}(b), one can find the response behaviours from two radical pairs are identical with weaker responses in a narrower response regime.
According to Fig.\ref{FIG:4-RP_couple_23_0} and Fig.\ref{FIG:4-RP_couple_23_1}, the responses are associated with the connected spin. The sensitive regime is narrower when the connection is established at $S_A$, broader when the connection is at $S_B$.

\subsection{The enabled sensing regime affected by local interaction}

\begin{figure}[h!]
\centering
\includegraphics[width=1\textwidth]{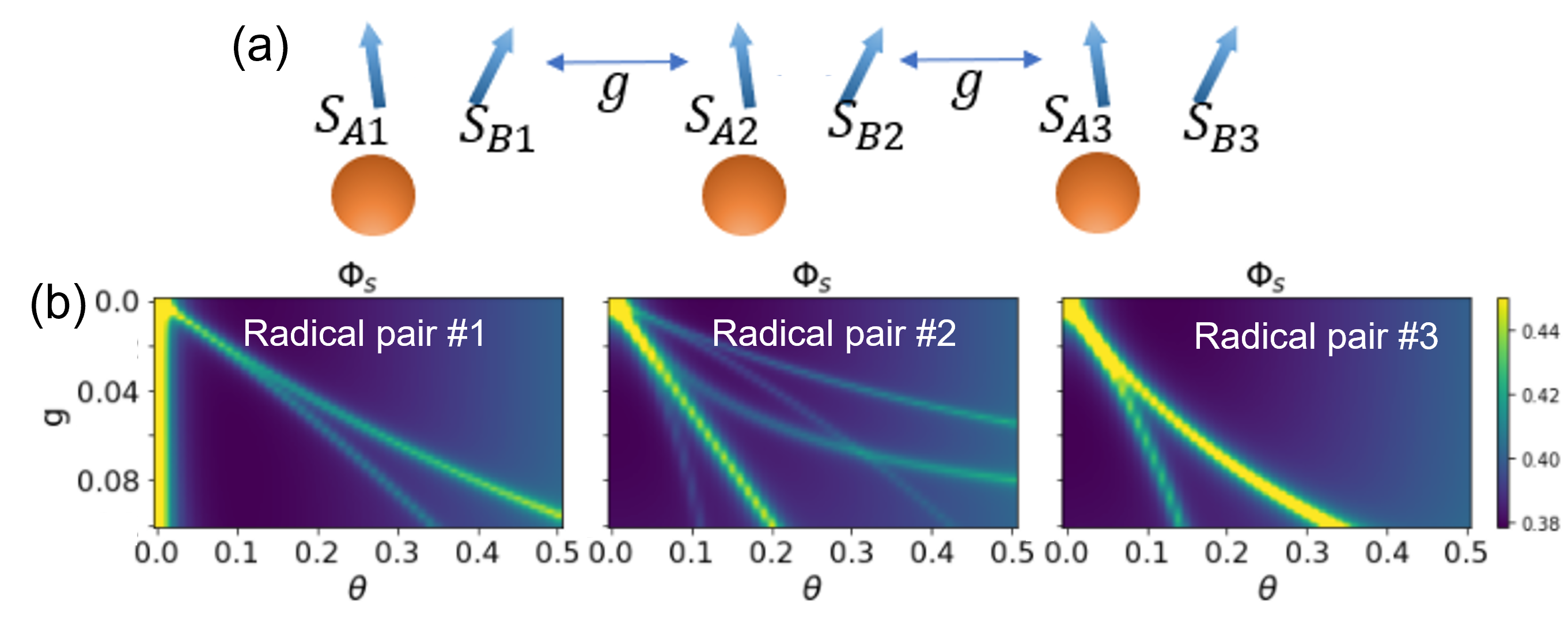}
\caption{(a) is the schematic representation of three-coupled radical pair system under the coupling of $G_2$. (b) is the contour plots of singlet yields under different coupling strength and magnetic strength.}
\label{FIG:5-RP_couple_1}
\end{figure}

\begin{figure}[h!]
\centering
\includegraphics[width=1\textwidth]{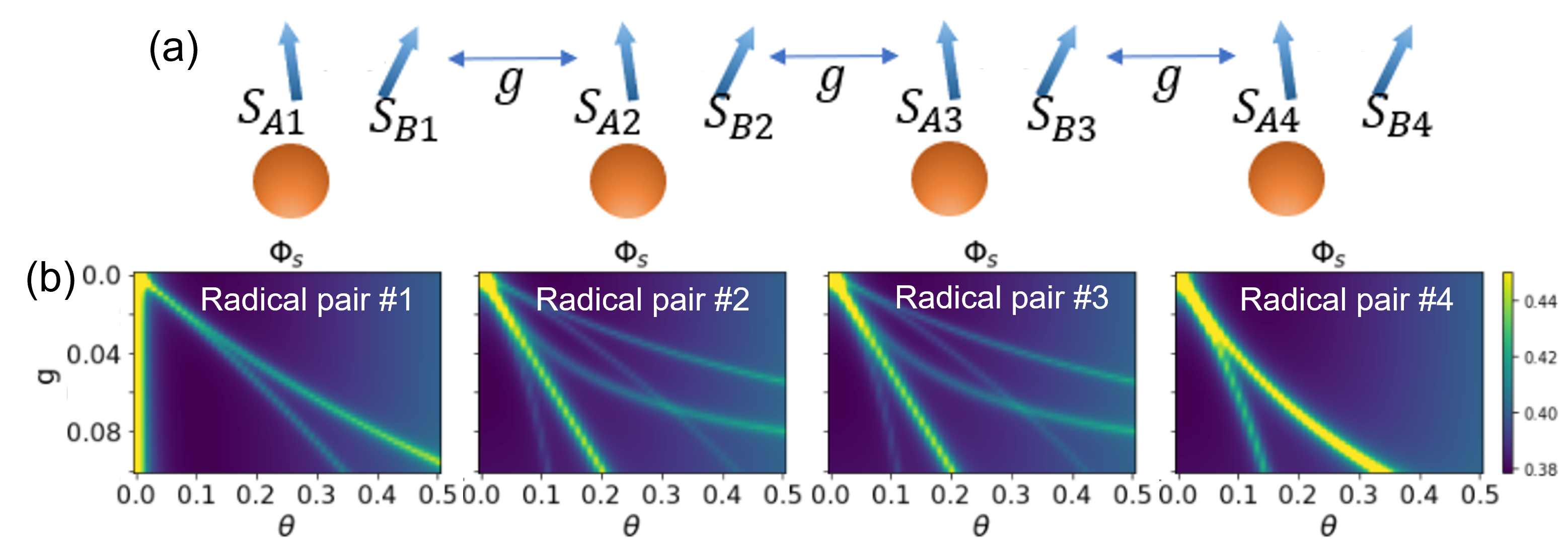}
\caption{(a) is the schematic representation of four-coupled radical pair system under the coupling of $G_2$. (b) is the contour plots of singlet yields under different coupling strength and magnetic strength.}
\label{FIG:5-RP_couple_2}
\end{figure}

System size and connection topology are critical in the complex system, such as phase transition and criticality. In metrology, increases in qubit number will improve the sensitivity due to the qubit interaction. To investigate the effect of network interaction, coupled system is extended to three- and four- coupled radical pair shown in Fig.\ref{FIG:5-RP_couple_1}(a) and Fig.\ref{FIG:5-RP_couple_2}(a) under $G_2$ coupling. Fig.\ref{FIG:5-RP_couple_1}(b) and Fig.\ref{FIG:5-RP_couple_2}(b) shows complicated response properties. According to the figures, we found the responses of the radical pairs located at the edges display the properties identical to the observation in two-coupled radical pair system under the $G_2$ coupling in Fig.\ref{FIG:4-RP_couple_23_0}, agreeing to the conjecture that the wider response regime for magnetic sensing when the coupling is happening at $S_B$ mentioned before. For the radical pair located at the middle, when $g$ is fixed, more peaks appear when the magnetic field is increased. The response of the system can be further quantified by sensitivity defined in the model section. We find the response to the external field is weaker and the sensitive regime is smaller, implying the radical pair at the middle may not be a good candidate for being the magnetic sensor. From the observation in Fig.\ref{FIG:5-RP_couple_1}(b) and Fig.\ref{FIG:5-RP_couple_2}(b), we find the response is only associated with the coupled spin, therefore we conjecture that, for $N$- coupled radical pair system, radical pair \#1 and \#$N$ show identical behaviour when $N \geq 2$ in different system size. 

\subsection{The initial state with quantum interference property shows better sensitivity in radical pair system}

In quantum sensing, it has been shown that the quantum advantage can be observed by introducing the entangled state as an initial condition. To confirm whether this can still be found in radical pair system, we consider the magnetic sensing scenario under two types of initial conditions; classical and non-classical state. ``Classical state'' is the initial condition prepared in a definite state. While the non-classical state refers the initial state 
is in quantum interference state. The density matrix of classical state $|C \rangle \langle C|$ is represented by $\frac{1}{2}(|01 \rangle \langle 01| + |10 \rangle \langle 10|)$ while the ``non-classica'' initial condition gives density matrix $|S\rangle \langle S|$(singlet superposition) and $|U\rangle \langle U|$(plus superposition), where $|S\rangle = \frac{1}{\sqrt{2}} (|01\rangle - |01\rangle)$, $|U\rangle = \frac{1}{2} (|00\rangle +|01\rangle+|10\rangle +|11\rangle)$. By applying different initial states to the isolated radical pair, Eq.\ref{totaly} gives the response curves and can be written as,

\begin{equation}
\begin{split}
\label{eqn:singlet_GAB0}
\Phi_s^c &= \frac{3}{8}+\frac{1}{8}\frac{\theta^2}{\Omega^2}+\frac{1}{8}\frac{a^2}{\Omega^2}f(\theta_1)\\ 
\Phi_s^s &= \frac{3}{8}+\frac{1}{8}\frac{\theta^2}{\Omega^2}+\frac{1}{8}\frac{a^2}{\Omega^2}f(\theta_1)+\frac{1}{8}(1-\frac{\theta^2}{\Omega^2})[f(\theta_2)+f(\theta_3)]+ \frac{1}{8}(1+\frac{\theta^2}{\Omega^2})[f(\theta_4) + f(\theta_5)]\\
\Phi_s^u &= \frac{1}{4}-\frac{1}{16}(1-\frac{\theta^2}{\Omega^2})[f(\theta_2)+f(\theta_3)]- \frac{1}{16}(1+\frac{\theta^2}{\Omega^2})[f(\theta_4)+f(\theta_5)]
\end{split}
\end{equation}

where $\Phi_s^c$, $\Phi_s^s$
and $\Phi_s^u$ are the singlet yields under the initial condition of ``classical'', ``singlet superposition'' and ``plus superposition'' state, respectively,
$\theta_1=\Omega$,
$\theta_2=\frac{a}{2}+\frac{\theta}{2}+\frac{\Omega}{2}$,
$\theta_3=\frac{a}{2}-\frac{\theta}{2}-\frac{\Omega}{2}$,
$\theta_4=\frac{a}{2}+\frac{\theta}{2}-\frac{\Omega}{2}$,
$\theta_5=\frac{a}{2}-\frac{\theta}{2}+\frac{\Omega}{2}$, $\Omega = \sqrt{1+\theta^2}$.

\begin{figure}[h!]
\centering
\includegraphics[width=0.5\textwidth]{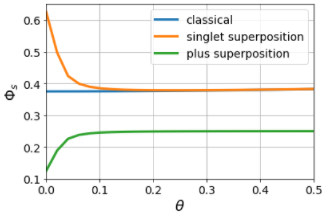}
\caption{The response curves with initial condition of classical, singlet superposition and plus superposition initial state}\label{FIG:CSP}
\end{figure}



From Fig.\ref{FIG:CSP} we can see the response is insensitive when the initial condition is at the classical state, while the sensitivities are compatible for singlet superposition and plus superposition state. The initial state is crucial in this sensing system as the yield production has been determined by Eq.\ref{totaly}, in which the projection of the initial state on the eigenspace of Hamiltonian plays an important role in the composition of singlet yield production. By applying different initial conditions to the system, we find the sensitivity is dependent on the initial state and conclude the classical initial condition(mixed state) has worse performance on sensitivity. For the system with the initial condition at quantum interference state, the entanglement state(singlet superposition) outperform the plus state at a small recombination rate.



\section{Discussion}

Magnetic field is ubiquitous in nature, whose sources range from neurons in the brain, current in circuits to the geodynamo at the core of the earth. Thus accurate and flexible magnetic sensing is of fundamental importance to multidisciplinary applications and research. Several sensing mechanisms have been uncovered to fulfil the required efficiency, cost, and accuracy for a vast range of measurement scenarios. However, the implementation of those mechanisms normally poses difficulty to control and lack flexibility. In this manuscript, we propose a quantum spin system consisting of coupled radical pairs to fulfil both the flexibility and accuracy requirements to sense the magnetic field. 

The radical pair system, such as the spins in \textit{Cryptochrome} or biradical pair molecule, is the stable system one can find in nature. Its recombination rate is also a key sensing parameter. We have observed the sensitivities decrease with recombination rate, $K$. Under the weak field approximation, the response is insensitive when the initial condition is at the classical state while the sensitivities of singlet superposition and mixture state are compatible with each other. We also found the critical $k^*$ that, the system with singlet initial condition outperforms the one with the plus superposition state when $k<K^*$. 

The quantum advantage results from coupled system are also investigated. Here we find increasing connectivity and introducing the full entangled initial state does not improve the sensitivity. Similar to the isolated radical pair system, the coupled system with the classical initial state has weak sensitivity, while the full entangled GHZ state does not show the advantage for sensing by comparing it to the singlet initial condition. This can be explained by; the singlet yield of radical pair system is expressed by the product of three terms, (i)the projection to the singlet state, (ii) the initial density matrix to the eigenstates of Hamiltonian and the (iii) accumulated transition within the spin lifetime. According to this analytic expression to the singlet yield, the first and third terms are not affected by changing the initial state, only the second term explicitly involves the initial state. Since the eigenstates remain the same, the initial density matrix that involves more combination states contributes more terms in the summation, giving a higher singlet yield level. In this manuscript, we only consider the local measurement to a specific radical pair within the system. The quantum advantage that can be achieved in a more sophisticated way involves the global phase measurement and is beyond the scope of this study needs to be investigated further. 

\begin{figure}[h!]
\centering
\includegraphics[width=0.5\textwidth]{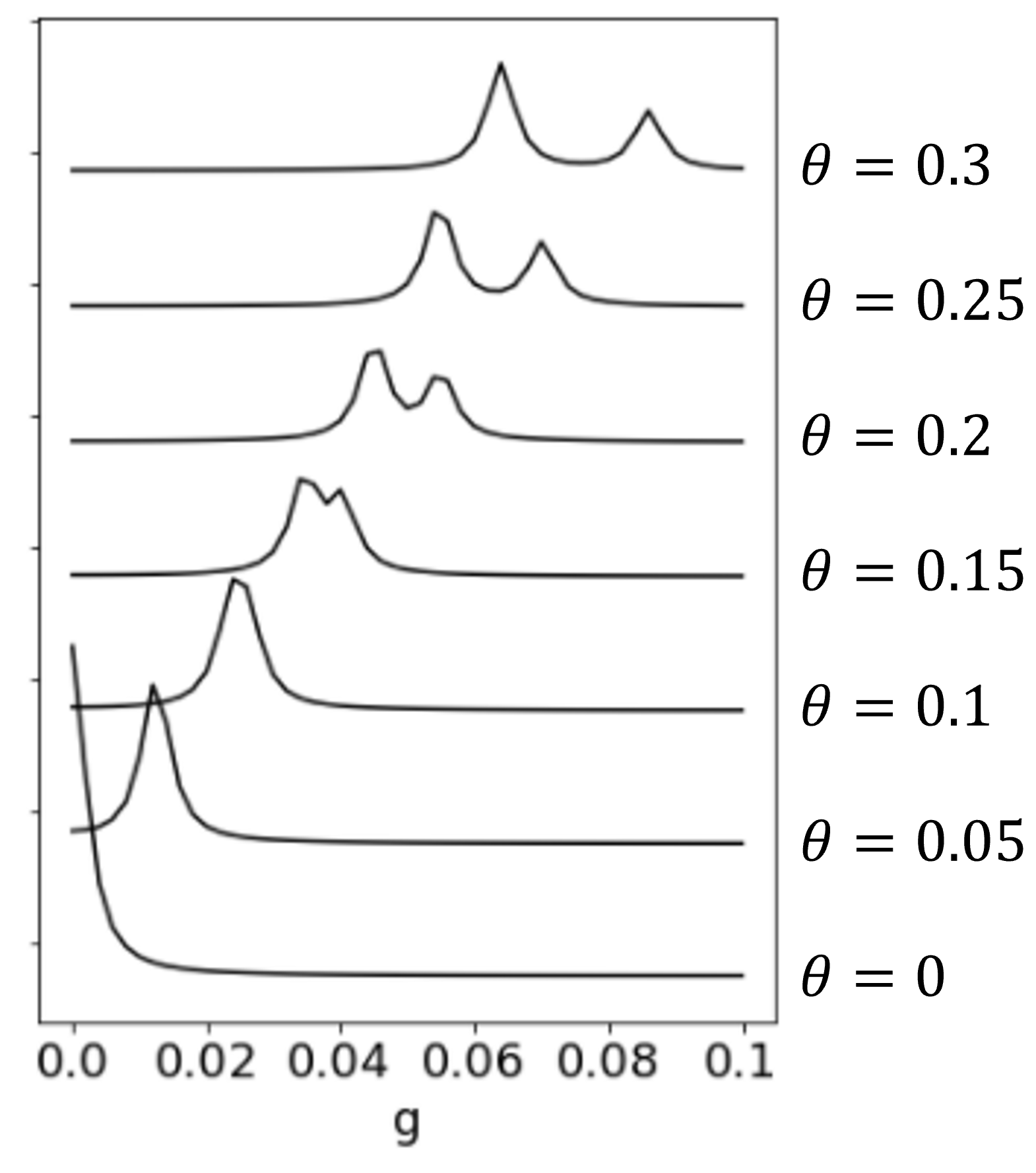}
\caption{Singlet yield under different magnetic field and coupling strength.}\label{FIG:radical_pair}
\end{figure}

In our proposed sensing method, the magnetic strength is inferred from the measured peak position in Fig.\ref{FIG:radical_pair} which is similar to the ESR spectrum of NV$^-$-defect. This is realized by degeneracy which results from the coupling strength and the intrinsic Hamiltonian under the external field. The sensing mechanism is similar to the NV$^{-}$-defect system, in which, the spin-selective process allows one to detect magnetic field strength by introducing a microwave to the system. In NV$^{-}$-defect system the resonance absorption appeared when the energy from microwave compatible to the energy gap between $m_s=0$ and $m_s= \pm 1$ under Zeeman effect at the ground state.

The spins are initialized by $m_s=0$ at the ground state. When absence of resonance, electrons are pumped to their excited state, then relax back accompanied by red light emission. However, the resonance can be raised by adjusting microwave to the appropriate frequency that corresponds to the energy gap. Electrons transit to the state of $m_s=\pm1$ from the ground state results in the non-radiative process, photon illumination intensity decreased. This resonance property again allows us to deduct the magnetic strength from the resonance frequency. In both coupled radial pair and NV$^{-}$-defect systems, the information of the magnetic field is encoded by Hamiltonian through the Zeeman effect. Resonances appear once the tunning parameter, i.e., coupling strength in coupled radical pair system, microwave frequency in NV defect, is compatible with the Hamiltonian under an external field. These resonances result in abrupt changes in observable quantities, enabling us to subtract the information from the magnetic field.

Although the sensing mechanism is similar, more flexibility can be found in coupled-radical pair systems. Because the biradical molecule size is in the order of nm, the spatial resolution of the radical pair will outperform the NV$^{-}$-defect as the spin location of the radical pair is controllable. The possible way for controlling the radical pair is, to attach biradical pair molecules to the beads which are controlled by optical tweezers. Therefore the distance between radical pairs can be adjusted by lasers. Based on this potential protocol, the scanning progress can be performed by adjusting the distance between molecules in the scale $<$ nm to modify coupling strength $g$. According to the estimation in our model by considering hyperfine interaction is around $0.01-10mT$, the sensed magnetic field can be in $\mu T$. Here we provide the concept that utilizes that spin dynamics to sense the magnetic field to have better spatial resolution and flexibility. The effect of buffer and the noise need to be further discussed.


\section*{Summary}
In this manuscript, we demonstrate magnetic sensing by showing the response pattern of simplified radical pair. We find the working regime for magnetic sensing varies with in intra-radical coupling $G_{AB}$. Studied in coupled radical pair shows collective sensing enables a more general and flexible sensing mechanism. The mechanism can be explained from the expression of singlet yield production, the energy spacing of coupled radical pair system is modulated by external magnetic field through the Zeeman effect. Changes in coupling strength between radical pairs brings a new parameter for energy spacing controlling. Under the appropriated coupling strength, energy spacing is squeezed, a high level of singlet yields can be produced when the initial state composed of interference state. Here we demonstrate the possibility of radical pair in biradical molecule can be used for magnetic sensing. Although the sensing mechanism shows the similarity to the NV$^{-}$-defect system, we hope this study can bring out a new type of measurement method which is flexible without losing sensitivity.



\printbibliography

\clearpage
\section*{Appendix}

We include more detail of the results mentioned in discussion in this section.

\subsection*{Effect of recombination on sensitivity}

\begin{figure}[h!]
\centering
\includegraphics[width=0.9\textwidth]{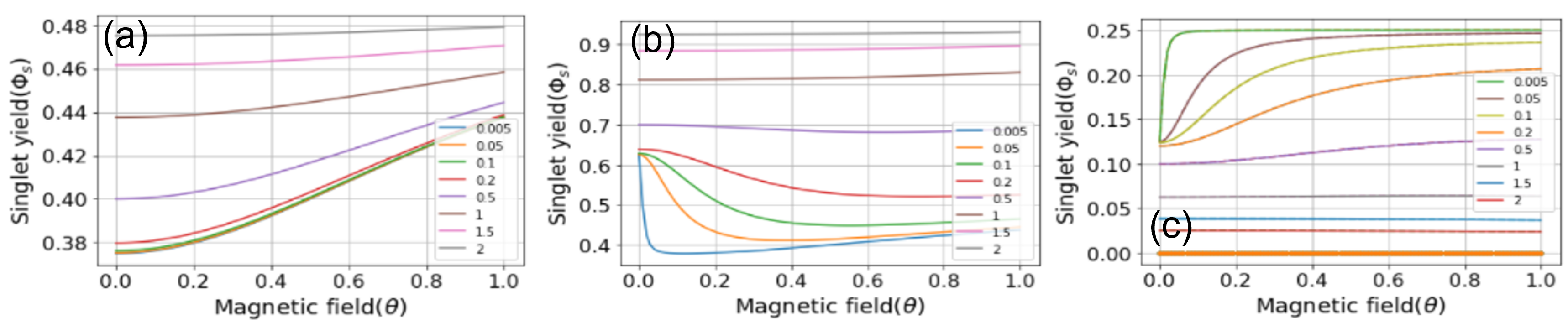}
\caption{The response curves under (a)classical (b)singlet superposition(c)plus superposition initial state with different recombination rate}\label{FIG:1_radical_diff_init}
\end{figure}

In the simplified radical pair model under the weak field approximation, we find the sensitivity of the system with different initial states can outperform one the other under different recombination rates. In Fig.\ref{FIG:1_radical_diff_init}, the response curves under (a)classical (b)singlet superposition(c)plus superposition initial state with different recombination rate $K$ show the singlet yield has strong dependence on the external field at low recombination rate, implies effective sensing appears when life time of radical pair is long.   

\begin{figure}[h!]
\centering
\includegraphics[width=0.9\textwidth]{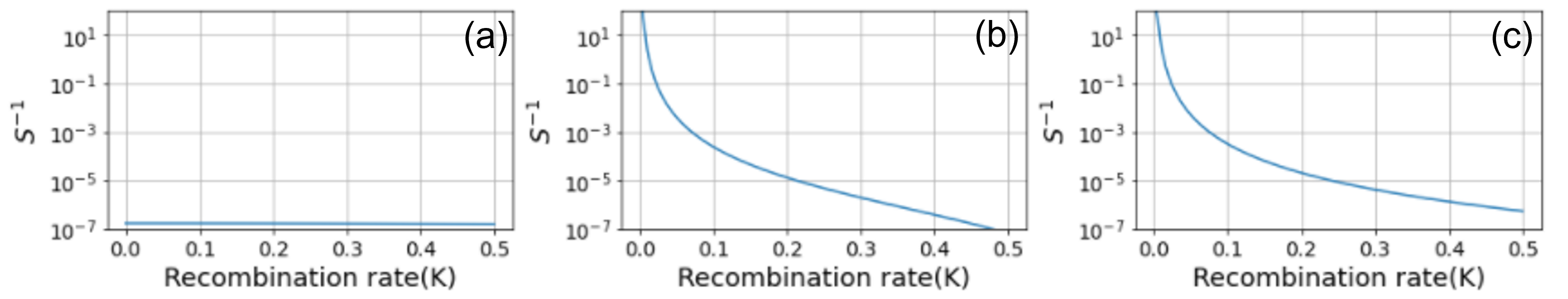}
\caption{The inverse of sensitivity curves under (a)classical (b)singlet superposition(c)plus superposition initial state with different recombination rate under weak field approximation.}\label{FIG:1_radical_diff_init_S}
\end{figure}

We have observed the better sensitivities in small recombination rate. It can be further quantified by $S$ through the weak field approximation. Through the approximation, Fig.\ref{FIG:1_radical_diff_init_S} shows the response is insensitive when the initial condition is at the classical state while the sensitivities are compatible for singlet superposition and plus superposition state. 

The difference between the sensitivity of quantum interference state can be identified by the ratio of sensitivity, $R$,\\
$$R(k) = \frac{S_s^{-1}}{S_u^{-1}} = \frac{2k^2a^2+\frac{1}{2}k^2\theta^2+a^2\theta^2}{4k^2+\frac{5}{2}k^2a^2+\frac{3}{4}k^2\theta^2+\frac{3}{8}a^2\theta^2} $$
Their sensitivities are equivalent when $R(K^*)=1$, therefore, 
$$K^* = [\frac{-(\frac{a^2}{2}+\frac{\theta^2}{4})\pm \sqrt{(\frac{a^2}{2}+\frac{\theta^2}{4})^2-10a^2\theta^2}}{8}]^{1/2}$$
From the expression, we can see when $k<K^*$, the system with singlet initial condition outperforms the one with the plus superpoisition state.

\subsection*{The benefit of sensing from coupling is not pronounced}
\begin{figure}[h!]
\centering
\includegraphics[width=0.9\textwidth]{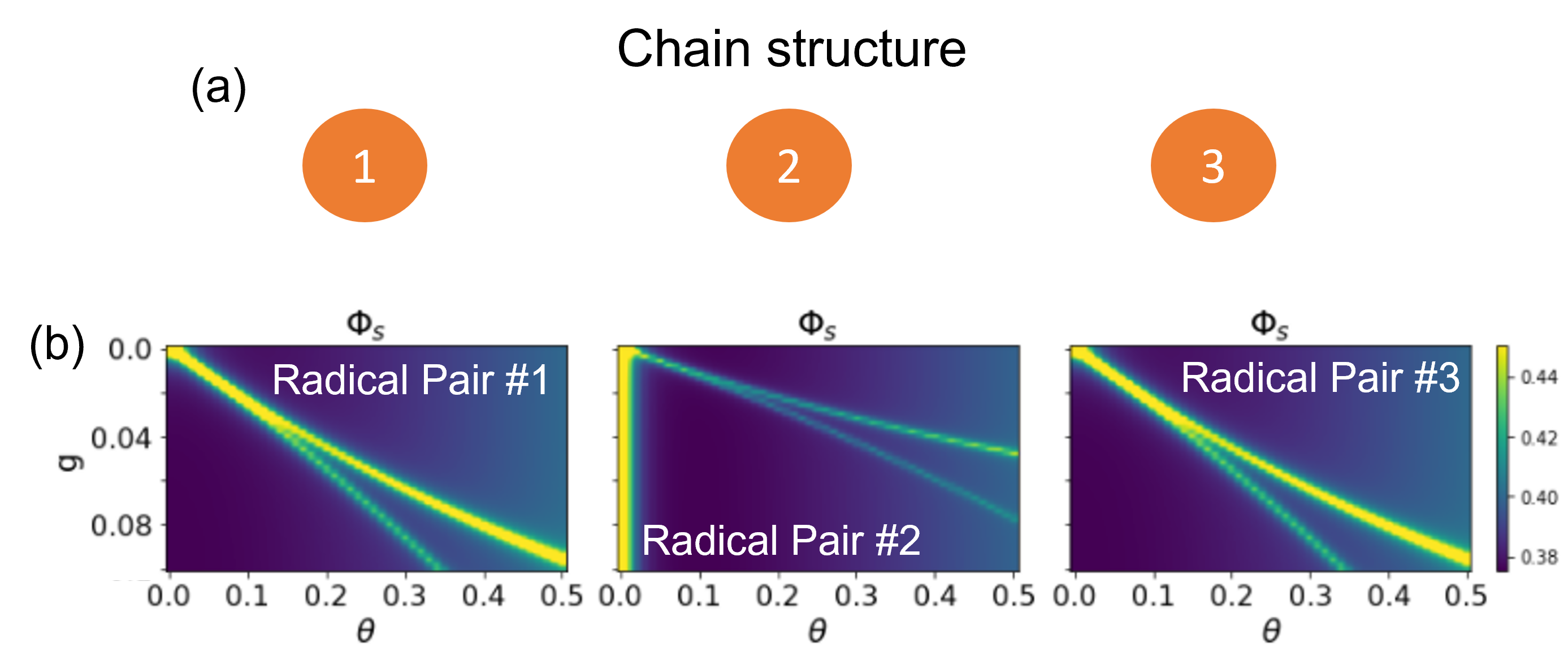}
\caption{(a)Three radical pairs are aligned under $G_4$ coupling. (b)The response pattern of one radical pair couple to another two radical pairs are calculated. In this three-coupled system, the response curve at the edge show the behavior that is identical to the two-coupled radical pair system. However, the radical pair located in the middle shows weaker response with wider sensing regime under the same coupling strength. }\label{FIG:3-RP_couple_chain}
\end{figure}

\begin{figure}[h!]
\centering
\includegraphics[width=0.9\textwidth]{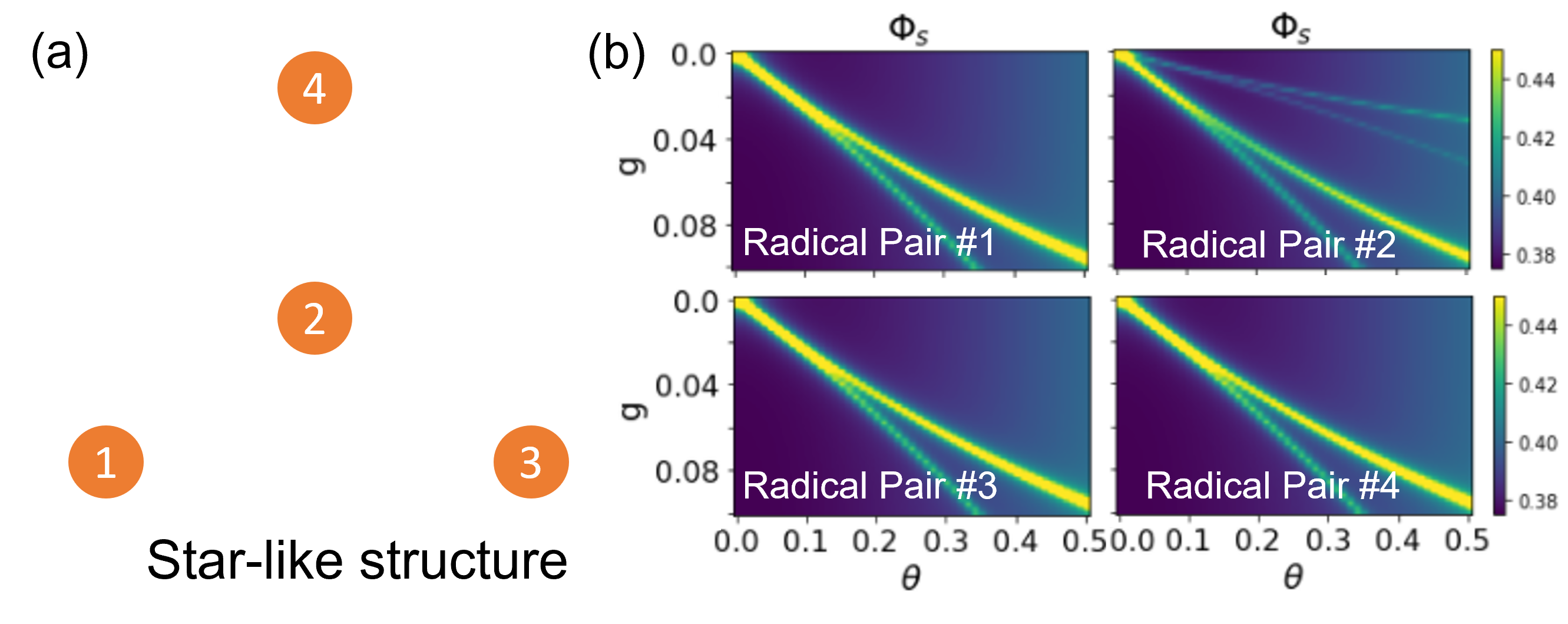}
\caption{(a)Four radical pairs in star-like topology under $G_4$ coupling is considered. (b)The response pattern of one radical pair couple to another three radical pairs are calculated. Addition connectivity makes the response pattern even more complicated by having more peaks.}\label{FIG:4-RP_couple_star}
\end{figure}

To identify the effect of different number of neighbors, we consider radical pairs aligned in the chain and star-like structure under $G_4$ coupling. From Fig. \ref{FIG:3-RP_couple_chain} and Fig.\ref{FIG:4-RP_couple_star}, we found the radical pair located at the edges shows the behaviors identical to the responses found at the edge node in two-coupled radical pair system while the response of the center radical pair becomes different in chain and star-like structure. When the radical pair with two neighbors, the response pattern becomes weaker and shifts to the small coupling regime. For the radical pair with three neighbors, the pattern becomes complicated. One can find its response pattern not only inherit the property from the node \#2 in the chain structure, but also the property  found at the edge nodes. From the observation, we found increases in connectivity will result in energy interference and not improving sensitivity.

\subsection*{Full entangled state does not further improve sensitivity}

\begin{figure}[h!]
\centering
\includegraphics[width=0.9\textwidth]{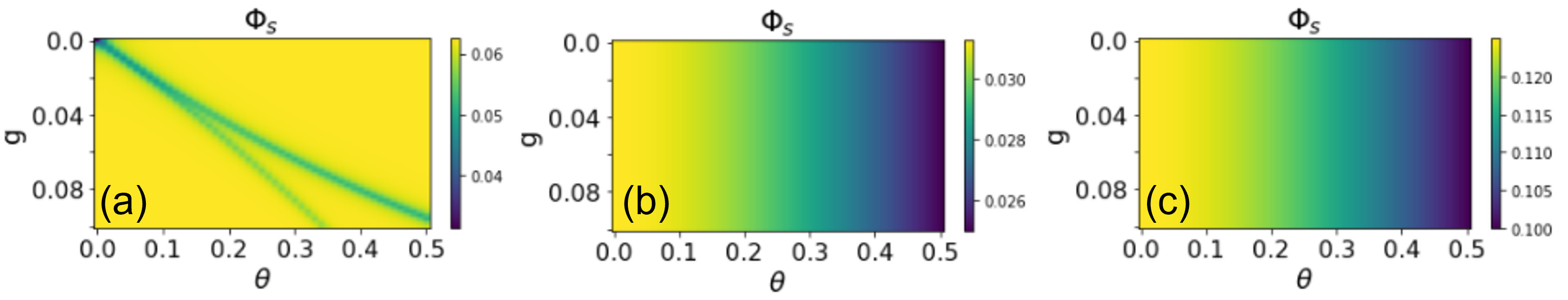}
\caption{Different initial conditions for 2-coupled radical pair system are shown. They are (a)$[\frac{1}{2}I_2 \otimes |u\rangle \langle u|]^{\otimes 2}$ with $|u\rangle= \frac{1}{2}(| 00 \rangle+ | 01 \rangle + | 10 \rangle + | 11 \rangle |)$ (b)classical GHZ state $[\frac{1}{2}I_2 \otimes \frac{1}{2} (|00\rangle \langle 00| + |11\rangle \langle 11| ]^{\otimes 2}$ (c)GHZ state state $[\frac{1}{2}I_2\otimes \frac{1}{4}(|00\rangle \langle 00|+|00\rangle \langle 11|+|11\rangle \langle00|+|11\rangle\langle 11|)]^{\otimes 2}$, respectively.  Fully entangled state does not improve the sensitivity by either increasing the response amplitude or creating the new sensing mechanism.}\label{FIG:8-RP_couple_initial_change}
\end{figure}

Entangled initial condition provided the quantum advantage on sensing accuracy. We investigate the advantage on radical pair system by introducing different initial states. The response patterns of singlet yield under the different initial conditions with $G_4$ coupling in coupled two-radical pair system and are shown in Fig.\ref{FIG:8-RP_couple_initial_change}. The initial condition with plus superposition state described by $[\frac{1}{2}I_2 \otimes |u\rangle \langle u|]^{\otimes 2}$ is considered. Where $|u\rangle= \frac{1}{2}(| 00 \rangle+ | 01 \rangle + | 10 \rangle + | 11 \rangle |)$. There is no entanglement in this initial state because the state of one spin does not correlate to any specific state of the other spin. Fig.\ref{FIG:8-RP_couple_initial_change}(a) shows the response patterns of two identical radical pairs. The patterns are identical to Fig.\ref{FIG:3-RP_couple_1}(b) except the peaks becomes canyons. 

The initial states with the classical GHZ state $[\frac{1}{2}I_2 \otimes \frac{1}{2} (|00\rangle \langle 00| + |11\rangle \langle 11|]^{\otimes 2}$ and GHZ state $[\frac{1}{2}I_2\otimes \frac{1}{4}(|00\rangle \langle 00|+|00\rangle \langle 11|+|11\rangle \langle00|+|11\rangle\langle 11|)]^{\otimes 2}$ are consider in Fig.\ref{FIG:8-RP_couple_initial_change}(b) and Fig.\ref{FIG:8-RP_couple_initial_change}(c). We found the response of this coupled system becomes monotonous, no peaks can be found in the system. Similar to the isolated radical pair system, the coupled system with the classical initial state has weak sensitivity, while the full entangled GHZ state does not show the advantage for sensing by comparing to the singlet initial condition. 

\clearpage

\section*{Response pattern when $G_{AB}= -0.1$}
\begin{figure}[h!]
\centering
\includegraphics[width=0.8\textwidth]{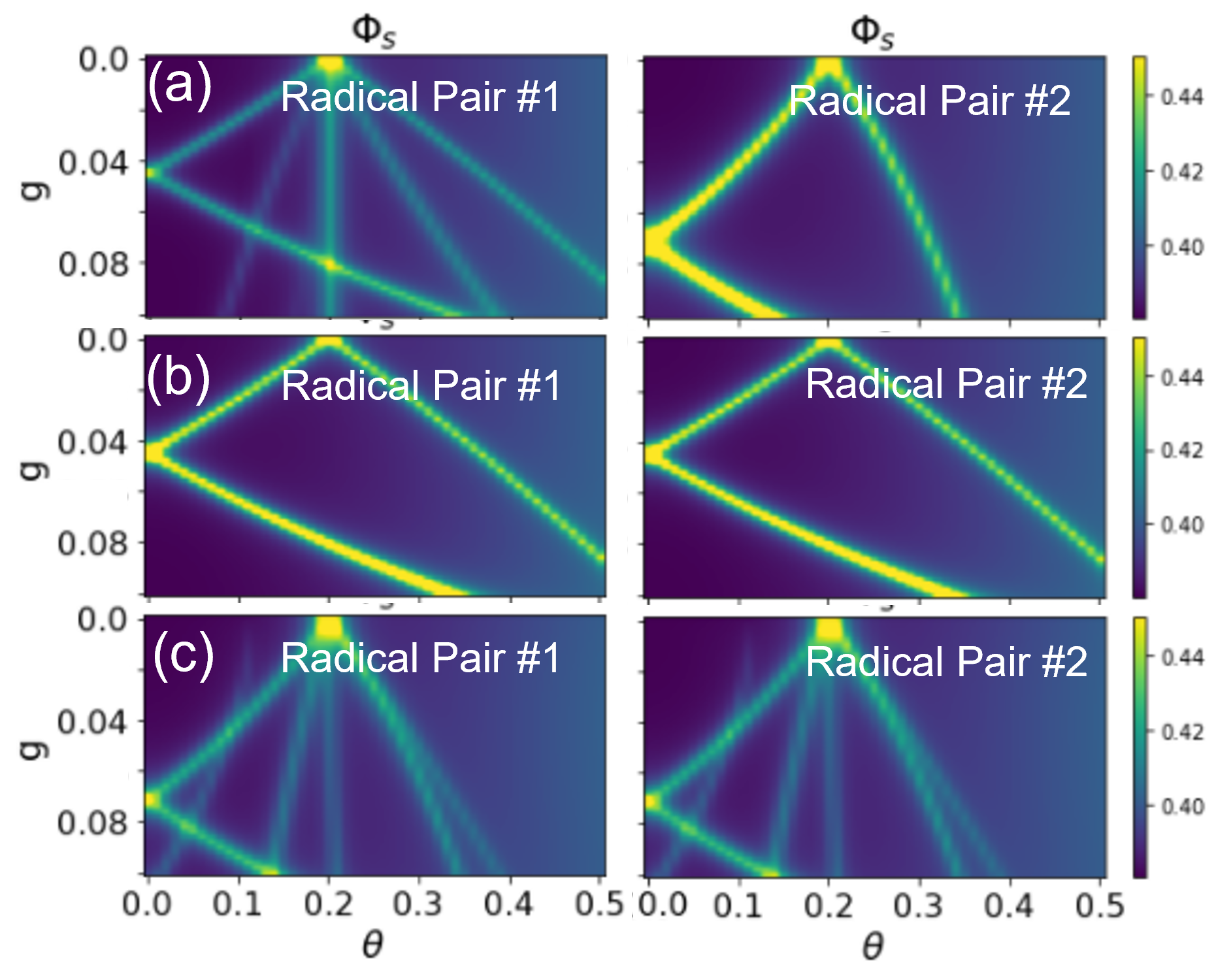}
\caption{Contour plot of singlet yield in both radical pairs under (a)$G_2$, (b)$G_4$ and (c)$G_1$ when coupling $G_{AB} = 0.1$. For $G_2$ coupling, radical pair\#1 shows the weaker response compare to $G_4$ while the response regime of radical pair \#2 is squeezed to the weaker magnetic regime.  Under $G_4$ coupling, symmetric configuration allows both radical pair \#1 and \#2 show the identical response to the external field under coupling. Under the $G_1$ coupling, complicated response pattern formed. Not only the peak splitting appears in small magnetic regime, but the response regime is squeezed compare to $G_2$.
}\label{FIG:G_AB_ALL}
\end{figure}

We demonstrate the results from two-coupled radical pair by considering the effect of dipolar interaction] and find the fundamental properties of response patterns are similar to the case with $G_{AB} = 0$ and will retain the conclusion shown in the results section.
The only difference is the sensitive regime becomes different while the topological dependency is the same. For example, from Fig.\ref{FIG:3-RP_couple_1}, Fig.\ref{FIG:4-RP_couple_23_0} and Fig.\ref{FIG:G_AB_ALL}, the response patterns of radical pair \#1 are the same under coupling of $\hat{S}_{B1}$-$\hat{S}_{A2}$($G_2$) and $\hat{S}_{B1}$-$\hat{S}_{B2}$($G_4$). Whether $G_{AB}$ is equal to zero does not change this conclusion. To utilize the radical pair as a new magnetic sensing scheme, the analytical expression for eigenvalue degeneracy of the new Hamiltonian need to be further investigated.

\clearpage

\end{document}